\documentclass[english]{sbrt}
\usepackage[english]{babel}
\usepackage[utf8]{inputenc}

\usepackage{cite}
\usepackage{float}
\usepackage{amsmath, amsfonts}
\usepackage[caption=false]{subfig}
\usepackage{graphicx}
\usepackage{multirow}

\usepackage[nolist,printonlyused]{acronym}

\begin{acronym}
	\acro{AP}{access point}
	\acro{CDF}{cumulative distribution function}
	\acro{CF-mMIMO}{Cell-free massive multiple-input multiple-output}
	\acro{CPU}{central processing unit}
	\acro{CSI}{channel state information}
	\acro{DL}{downlink}
	\acro{FR3}{frequency range 3}
	\acro{MMSE}{minimum mean square error}
	\acro{MR}{maximum ratio}
	\acro{P-AP}{primary AP}
	\acro{P-UE}{primary UE}
	\acro{PN}{primary network}
	\acro{QoS}{quality of service}
	\acro{S-AP}{secondary AP}
	\acro{S-UE}{secondary UE}
	\acro{SE}{spectral efficiency}
	\acro{SINR}{signal-to-interference-plus-noise ratio}
	\acro{SN}{secondary network}
	\acro{SoS}{sum-of-sinusoids}
	\acro{TDD}{time division duplex}
	\acro{UE}{user equipment}
	\acro{UL}{uplink}
\end{acronym}

\newcommand{\SecRef}[2][]{Section#1~\ref{#2}}
\newcommand{\FigRef}[2][]{Figure#1~\ref{#2}}
\newcommand{\TabRef}[2][]{Table#1~\ref{#2}}

\newcommand{\pn}{\mathrm{pn}}
\newcommand{\sn}{\mathrm{sn}}

\newcommand{\idxsap}{l}   
\newcommand{\idxpap}{b}   

\newcommand{\idxsue}{k}   
\newcommand{\idxpue}{j}   

\newcommand{\antPN}{M}    
\newcommand{\antSN}{N}    

\newcommand{\numSAP}{L}   
\newcommand{\numPAP}{B}   

\newcommand{\numSUE}{K}   
\newcommand{\numPUE}{J}   

\newcommand{\chSN}{\mathbf{h}}      
\newcommand{\chSP}{\mathbf{u}}      
\newcommand{\chPS}{\mathbf{v}}      
\newcommand{\chPN}{\mathbf{f}}      

\newcommand{\RSN}{\mathbf{R}}      
\newcommand{\RSP}{\mathbf{U}}      
\newcommand{\RPS}{\mathbf{V}}      
\newcommand{\RPN}{\mathbf{D}}

\begin{document}

\title{Zone-Based Interference Management for Underlay Spectrum Sharing in Cell-Free Networks}

\author{Mariana M. Ferreira, Juno V. Saraiva, Yuri C. B. Silva, Francisco R. P. Cavalcanti, \\ Igor M. Guerreiro, Roberto P. Antonioli, and Behrooz Makki
\thanks{Mariana M. Ferreira, Juno V. Saraiva, Yuri C. B. Silva, Francisco R. P. Cavalcanti, Igor M. Guerreiro, and Roberto P. Antonioli, Wireless Telecom Research Group, Federal University of Ceará, Fortaleza-CE, e-mail: \{mariana, juno, yuri, rodrigo, igor, antonioli\}@gtel.ufc.br; Behrooz Makki, Ericsson Research, Gothenburg, Sweden, e-mail: behrooz.makki@ericsson.com. This work was supported in part by Ericsson Research, Sweden, and Ericsson Innovation Center, Brazil, Technical Cooperation Contracts UFC.53 and UFC.55, in part by the Brazilian National Council for Scientific and Technological Development (CNPq), in part by CNPq/INCT-Signals Grant 406517/2022-3, in part by CAPES - Finance Code 001.}%
}

\maketitle


\begin{abstract}
Underlay spectrum sharing enables flexible coexistence between a \ac{PN} and a \ac{SN}, but inter-network interference remains a critical limiting factor. Existing mitigation strategies often rely on instantaneous \ac{CSI}, incurring high signaling overhead. This work proposes a zoning-based interference management framework that leverages large-scale channel statistics to partition the \ac{PN} into interference zones and classify \acp{S-AP} into risk tiers. The zone-level characterization enables selective power-reduction policies with reduced coordination complexity. Simulation results confirm that the proposed framework protects the \ac{PN} while preserving \ac{SN} \ac{SE} in distributed cell-free deployments.
\end{abstract}
\begin{keywords}
Spectrum Sharing, Underlay Coexistence, Cell-Free, Zone-Based Interference Management.
\end{keywords}

\acresetall

\section{Introduction} \label{section_introduction}

The continuous advancement of wireless connectivity and the rise of data-centric applications have intensified the demand for spectral resources. While the evolution toward wider bandwidths in the \ac{FR3} is expected to alleviate this limitation \cite{katwe2024}, spectrum scarcity continues to constrain the deployment of traditional exclusive-access networks. To overcome spectral limitations, spectrum-sharing paradigms have emerged as a fundamental approach for enabling the coexistence of different networks in the same frequency bands~\cite{Ahmad2020}.

\ac{CF-mMIMO}, where geographically distributed \acp{AP} cooperatively serve users across a shared area, has emerged as a promising architecture for shared-spectrum operation~\cite{Demir2021, Antonioli2022, Shaik2024}. Based on this architecture, we consider a spectrum-sharing scenario consisting of a \ac{PN}, which owns the spectrum license, and a \ac{SN} operating in underlay mode~\cite{Shaik2024}. In this setup, the \ac{SN} must strictly limit the interference imposed on the \ac{PN} while simultaneously satisfying its own \ac{QoS} requirements.

This regulation can be applied uniformly across all \acp{S-AP}, which is simple but unnecessarily penalizes nodes that contribute marginally to the aggregate interference. Alternatively, mitigation can be applied selectively, targeting only the \acp{S-AP} that pose the highest interference risk to the \ac{PN} \cite{Filho2025}. Realizing this selective approach, however, requires accurately identifying which nodes are interfering and to what extent.

Several studies have proposed interference cartography techniques to characterize the spatial distribution of interference, enabling the identification of restricted transmission regions and admissible power levels \cite{Sato2017, Naranjo2014, Ureten2012, Alaya2008}. Despite their potential, these approaches face substantial challenges: continuous spatial measurements are impractical in dynamic environments, and enforcing interference constraints often requires instantaneous \ac{CSI} exchange and cross-network coordination, resulting in excessive signaling overhead that limits scalability \cite{Hoyhtya2016}. Therefore, localized solutions that rely on long-term channel statistics are essential for practical coexistence.

To this end, this work proposes a zoning-based interference management framework for cell-free underlay spectrum sharing. By characterizing interference potential through large-scale fading statistics and abstracting it at the zone level, our proposed approach substantially reduces coordination overhead while preserving the essential spatial structure required for interference containment. Selective mitigation is then applied only to high-risk nodes, maintaining the overall utility of the \ac{SN} while protecting the \ac{PN}.

\section{System Model} \label{section_system_model}

A cell-free network consists of geographically distributed \acp{AP} jointly serving all \ac{UE} devices in the coverage area, connected to a \ac{CPU} through fronthaul links for cooperative transmission and reception~\cite{Demir2021}. We consider a spectrum-sharing scenario with two coexisting cell-free networks~\cite{Shaik2024}, a licensed \ac{PN} comprising $\numPAP$ \acp{P-AP} with $\antPN$ antennas each, jointly serving $\numPUE$ single-antenna \acp{P-UE}, and an \ac{SN} consisting of $\numSAP$ \acp{S-AP} with $\antSN$ antennas each, jointly serving $\numSUE$ single-antenna \acp{S-UE}, operating in underlay mode within the same frequency bands.

For each \ac{AP}-\ac{UE} pair, we define the channels: $\chSN_{\idxsue \idxsap} \in \mathbb{C}^{\antSN}$ from the $\idxsap$-th \ac{S-AP} to the $\idxsue$-th \ac{S-UE}, $\chSP_{\idxpue \idxsap} \in \mathbb{C}^{\antSN}$ from the $\idxsap$-th \ac{S-AP} to the $\idxpue$-th \ac{P-UE}, $\chPS_{\idxsue \idxpap} \in \mathbb{C}^{\antPN}$ from the $\idxpap$-th \ac{P-AP} to the $\idxsue$-th \ac{S-UE}, and $\chPN_{\idxpue \idxpap} \in \mathbb{C}^{\antPN}$ from the $\idxpap$-th \ac{P-AP} to the $\idxpue$-th \ac{P-UE}. All channels follow correlated Rayleigh fading: $\chSN_{\idxsue \idxsap} \sim \mathcal{CN}(\mathbf{0}, \RSN_{\idxsue \idxsap})$, $\chSP_{\idxpue \idxsap} \sim \mathcal{CN}(\mathbf{0}, \RSP_{\idxpue \idxsap})$, $\chPS_{\idxsue \idxpap} \sim \mathcal{CN}(\mathbf{0}, \RPS_{\idxsue \idxpap})$, and $\chPN_{\idxpue \idxpap} \sim \mathcal{CN}(\mathbf{0}, \RPN_{\idxpue \idxpap})$, where the spatial correlation matrices capture large-scale fading and spatial correlation effects.

Both networks operate under synchronized \ac{TDD}, where each coherence block of length $\tau_c$ is divided into $\tau_p$ pilots, $\tau_u$ \ac{UL} data, and $\tau_d$ \ac{DL} data symbols, such that $\tau_c=\tau_p+\tau_u+\tau_d$~\cite{Bjornson2020}. The pilot set is partitioned as $\boldsymbol{\Phi} = [\boldsymbol{\Phi}_s, \boldsymbol{\Phi}_0, \boldsymbol{\Phi}_p]$, where $\boldsymbol{\Phi}_s$ contains $\tau_1$ pilots exclusive to \acp{S-UE}, $\boldsymbol{\Phi}_0$ contains $\tau_2$ pilots shared between both networks, and $\boldsymbol{\Phi}_p$ contains $\tau_3$ pilots exclusive to \acp{P-UE}, with $\tau_p = \tau_1 + \tau_2 + \tau_3$~\cite{Shaik2024}.

\subsection{Channel Estimation}

During the \ac{UL} pilot phase, both \acp{P-UE} and \acp{S-UE} transmit their respective pilot sequences, and each \ac{AP} estimates the channels from its associated \acp{UE} using the \ac{MMSE} estimator described in~\cite{Shaik2024}. Assuming \ac{TDD} operation and channel reciprocity, the \ac{UL} estimates ($\hat{\chPN}_{\idxpue \idxpap}$ from the $\idxpap$-th \ac{P-AP} to the $\idxpue$-th \ac{P-UE} and $\hat{\chSN}_{\idxsue \idxsap}$ from the $\idxsap$-th \ac{S-AP} to the $\idxsue$-th \ac{S-UE}) are reused for \ac{DL} precoding without additional overhead.

\subsection{Downlink Data Transmission} 

In the \ac{DL}, each \ac{P-AP} and \ac{S-AP} constructs precoders $\mathbf{w}_{\idxpue \idxpap}^{\pn} \in \mathbb{C}^{\antPN}$ and $\mathbf{w}_{\idxsue \idxsap}^{\sn} \in \mathbb{C}^{\antSN}$, respectively. The transmit signal at the $\idxpap$-th \ac{P-AP} and the $\idxsap$-th \ac{S-AP} are
\begin{equation} \label{eq:transmit_signal}
	\mathbf{x}_{\idxpap}^{\pn} = \sum\limits_{\idxpue=1}^{\numPUE}\sqrt{ \rho_{\idxpue \idxpap}} \, \mathbf{w}_{\idxpue \idxpap}^{\pn} q_\idxpue, \quad \mathbf{x}_{\idxsap}^{\sn} = \sum\limits_{\idxsue=1}^{\numSUE}\sqrt{ \rho_{\idxsue \idxsap}} \mathbf{w}_{\idxsue \idxsap}^{\sn} q_\idxsue,
\end{equation}
\noindent where $q_{\idxpue}$ and $q_{\idxsue}$ are unit-power payload symbols, $\mathbf{w}_{\idxpue \idxpap}^{\pn}$ and $\mathbf{w}_{\idxsue \idxsap}^{\sn}$ have unit norm, and $\rho_{\idxpue \idxpap} \geq 0$, $\rho_{\idxsue \idxsap} \geq 0$ denotes the allocated transmit power. Each \ac{AP} is subject to a maximum transmit power constraint, such that $\sum_{\idxpue=1}^{\numPUE} \rho_{\idxpue \idxpap} \leq P_{\max, \idxpap}$ and $\sum_{\idxsue=1}^{\numSUE} \rho_{\idxsue \idxsap} \leq P_{\max, \idxsap}$ for each \ac{P-AP} and \ac{S-AP}, respectively.

The received signal at the $\idxpue$-th \ac{P-UE} is
\begin{equation}
	\begin{aligned}
		y_{\idxpue} &=\sum\limits_{\idxpap=1}^{\numPAP} \chPN_{\idxpue \idxpap}^{\mathrm{H}} \mathbf{x}_{\idxpap}^{\pn} 
        + \sum\limits_{\idxsap=1}^{\numSAP} \chSP_{\idxpue \idxsap}^{\mathrm{H}} \mathbf{x}_{\idxsap}^{\sn} + n_{\idxpue} \\ 
		&= \sum\limits_{\idxpap=1}^{\numPAP} \sqrt{\rho_{\idxpue \idxpap}} \chPN_{\idxpue \idxpap}^{\mathrm{H}} \mathbf{w}_{\idxpue \idxpap}^{\pn} q_{\idxpue} + \sum\limits_{i \neq \idxpue}^{\numPUE} \sum\limits_{\idxpap=1}^{\numPAP} \sqrt{\rho_{i \idxpap}} \chPN_{\idxpue \idxpap}^{\mathrm{H}} \mathbf{w}_{i \idxpap}^{\pn} q_i + n^{+}_{\idxpue},
		\label{eq:received_downlink_signals_pn}
	\end{aligned}
\end{equation}
\noindent where $n_{\idxpue}$ is additive noise with variance $\varsigma^2$, and $n^{+}_{\idxpue}$ denotes the inter-network interference-plus-noise term. 

\subsection{Performance Metrics}

The \ac{DL} \ac{SE} for the $\idxpue$-th \ac{P-UE} is
\begin{equation} \label{eq:spectral_efficiency_pn}
	\mathrm{SE}_{\idxpue} = \frac{\tau_d}{\tau_c} \log_2 (1 + \text{SINR}_{\idxpue}^{\pn}),
\end{equation}
\noindent where the \ac{SINR} is defined as 
\begin{equation}
    \text{SINR}_{\idxpue}^{\pn} = \frac{\left| \sum\limits_{\idxpap=1}^{\numPAP} \sqrt{\rho_{\idxpue \idxpap}} \chPN_{\idxpue \idxpap}^{\mathrm{H}} \mathbf{w}_{\idxpue \idxpap}^{\pn} \right|^2}{\sum\limits_{i \neq \idxpue}^{\numPUE} \left|\sum\limits_{\idxpap=1}^{\numPAP} \sqrt{\rho_{i \idxpap}} \chPN_{\idxpue \idxpap}^{\mathrm{H}} \mathbf{w}_{i \idxpap}^{\pn} \right|^2  + \left|\sum\limits_{\idxsap=1}^{\numSAP} \chSP_{\idxpue \idxsap}^{\mathrm{H}} \mathbf{x}_{\idxsap}^{\sn} \right|^2 + \varsigma^2}. 
\end{equation}

The \ac{DL} \ac{SE} for the \acp{S-UE} can be obtained following a procedure analogous to~\eqref{eq:spectral_efficiency_pn}. In underlay spectrum-sharing operation, cross-network interference affects both networks. However, since the \ac{PN} owns the spectrum license, the \ac{SN} must ensure that the aggregate interference power generated at each \ac{P-UE} remains below a predefined threshold~\cite{Alaya2008}, motivating the interference management framework proposed in \SecRef{section_zone}.

\section{Zone-Based Interference Management}\label{section_zone}

The cross-network interference is spatially non-uniform across the \ac{PN} coverage area, since \acp{P-UE} located near active \acp{S-AP} experience significantly stronger interference than those in distant regions and, reciprocally, \acp{S-AP} closer to the \ac{PN} emerge as dominant interference sources, so that in a side-by-side deployment the critical region concentrates at the network boundary. This motivates characterizing and mitigating interference at a regional granularity rather than through a uniform, network-wide policy, reducing coordination overhead and avoiding unnecessary power restrictions on low-risk nodes.

\begin{figure}[!h] 
    \centering
    \includegraphics[width=0.85\linewidth]{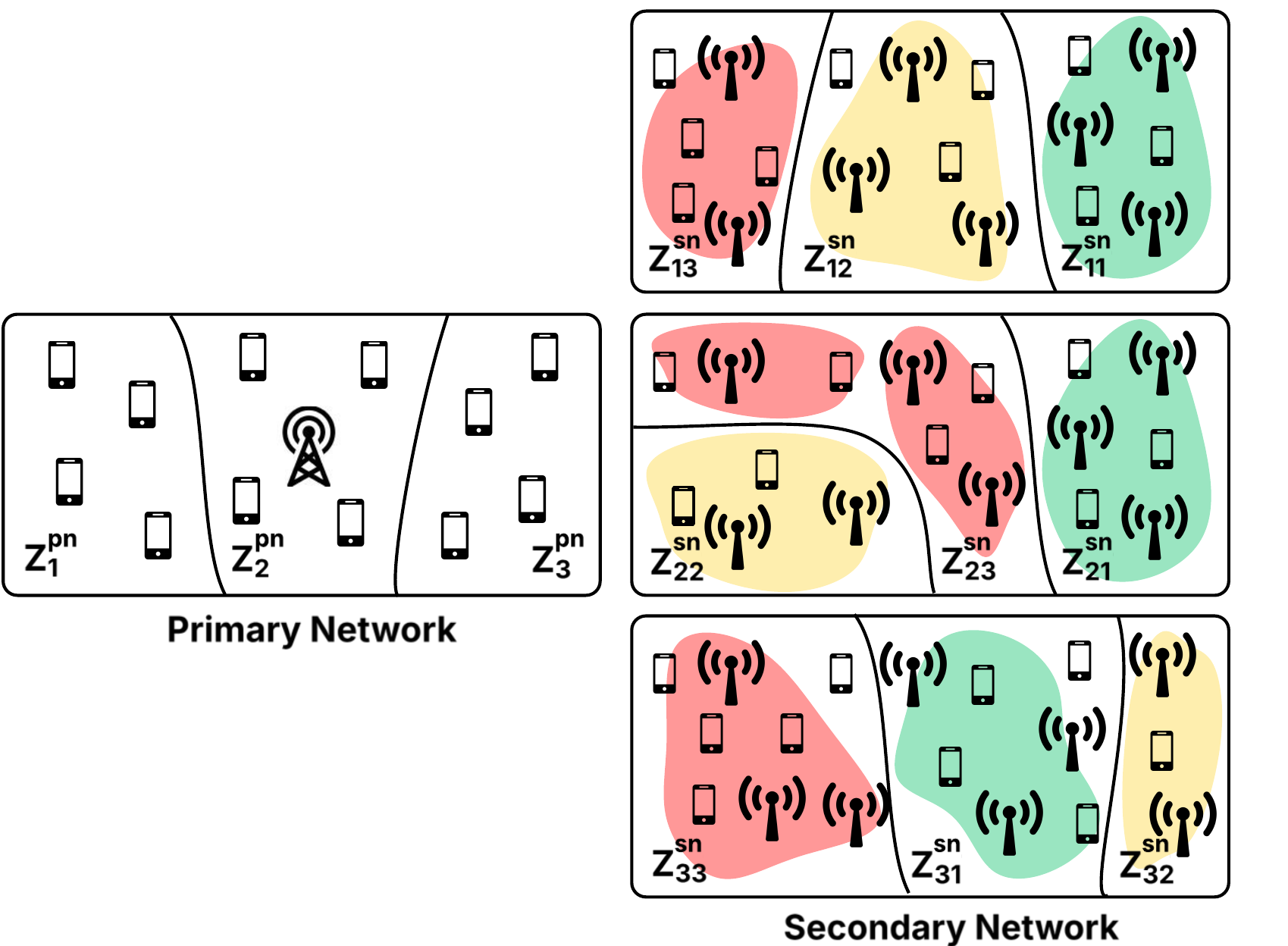}
    \caption{Proposed zoning method for inter-network interference management.}
    \label{fig:example_zones}
\end{figure}

\FigRef{fig:example_zones} illustrates this zone-based approach, where the \ac{PN} coverage area is partitioned into $N_z^{\pn} = 3$ zones, and for each \ac{PN} zone, the \acp{S-AP} are classified into $N_z^{\sn} = 3$ risk tiers. Green markers denote low-risk \acp{S-AP} with negligible impact on \acp{P-UE}, while yellow and red markers indicate medium and high-risk nodes, respectively.

\subsection{Large-Scale Interference Potential}

The zoning framework is grounded in large-scale channel statistics, which reflect long-term propagation characteristics without requiring instantaneous \ac{CSI} exchange. The interference potential of the $\idxsap$-th \ac{S-AP} at a spatial point $\mathbf{p}$ within the \ac{PN} coverage area is defined as
\begin{equation}
    I_{\idxsap}(\mathbf{p}) = P_{\max, \idxsap} \cdot \beta_{\idxsap}(\mathbf{p}),
	\label{eq:interference_potential}
\end{equation}
\noindent where $P_{\max, \idxsap}$ is the maximum transmit power and $\beta_{\idxsap}(\mathbf{p})$ denotes the large-scale channel gain between the $\idxsap$-th \ac{S-AP} and location $\mathbf{p}$. This metric jointly captures the effect of transmit power and propagation conditions, providing a long-term characterization of the interference risk posed by each \ac{S-AP} to any given point in the \ac{PN} coverage area.

\subsection{Primary Network Partitioning}

The \ac{PN} partitions its coverage area into $N_z^{\pn}$ zones based on the spatial distribution of the aggregate interference potential $I(\mathbf{p}) = \sum_{\idxsap=1}^{\numSAP} I_{\idxsap}(\mathbf{p})$, which captures the cumulative large-scale coupling from all \acp{S-AP} to each point $\mathbf{p}$ in the \ac{PN} coverage area. In practice, this partitioning can rely on interference reports collected from the \acp{P-UE} or, when only topology information is available, on the spatial distribution of the \acp{P-AP}.

Regardless of the partitioning strategy, a K-means algorithm is applied to the resulting values, with the number of zones $N_z^{\pn}$ either predefined or selected using the Silhouette Score~\cite{SilhouettesScore1987}, which ensures that the resulting zones are well-separated and internally cohesive. The \ac{PN} zones $Z_i^{\pn}$, $i = 1, \ldots, N_z^{\pn}$, are indexed in decreasing order of mean interference potential, so that $Z_1^{\pn}$ is the most exposed zone, enabling the \ac{PN} to enforce differentiated protection requirements across zones.

\subsection{Secondary Network Partitioning}

For each \ac{PN} zone $Z_i^{\pn}$, the \ac{SN} computes the spatial mean of $I_{\idxsap}(\mathbf{p})$, as defined in \eqref{eq:interference_potential}, over the sample points within $Z_i^{\pn}$ for every \ac{S-AP}. A one-dimensional K-means algorithm then partitions the \acp{S-AP} into $N_z^{\sn}$ risk tiers $Z_{it}^{\sn}$, $t = 1, \ldots, N_z^{\sn}$, ordered from lowest to highest interference level (e.g., low, medium, and high for $N_z^{\sn} = 3$), with $N_z^{\sn}$ predefined or selected via the Silhouette Score. Although the tiers are defined at the \ac{S-AP} level, each one has a spatial extent, obtained as the union of the Voronoi cells of the \acp{S-AP} it contains.

The zone classification described above directly enables selective interference mitigation. When the \ac{PN} identifies a zone $Z_i^{\pn}$ requiring protection, it notifies the \ac{SN} through a low-overhead signaling mechanism, ranging from a simple binary indicator (e.g., ``Zone $i$ is impaired'') to a more detailed mapping of the contributing risk tiers. Upon receiving this indication, the \ac{SN} applies mitigation, such as power reduction, only to \acp{S-AP} classified above a configurable risk tier.

\section{Simulation Results}\label{section_results}

\subsection{Simulation Setup}

\begin{figure}[h!]
	\centering
	\includegraphics[width=0.9\linewidth]{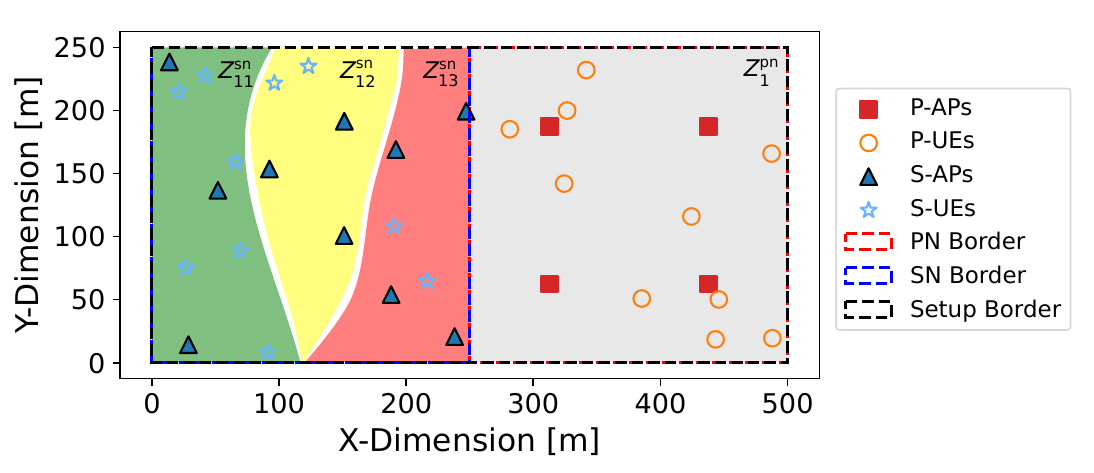}
	\caption{Coexistence scenario layout with $N_z^{\pn} = 1$ and $N_z^{\sn} = 3$: \ac{S-AP} risk classification into low- (green), medium- (yellow), and high-risk (red) tiers.}
	\label{fig:scenario_layout}
\end{figure}

The coexistence scenario considers both networks operating over a total area of $500~\text{m}\times250~\text{m}$, where each network occupies a $250~\text{m} \times 250~\text{m}$ region, as illustrated in \FigRef{fig:scenario_layout}.

Detailed topology parameters are summarized in \TabRef{table:network_topology}.
\begin{table}[H]
\centering
\caption{Network topology} 
\begin{tabular}{|c|c|c|}
    \hline
    \textbf{Parameter} & \textbf{Primary Network} & \textbf{Secondary Network}\\
    \hline\hline
    Number of \acp{AP} & $4$ & $10$ \\ \hline
    Antennas per \ac{AP} & $8$ & $4$ \\ \hline
    \ac{AP} distribution & Regular grid & Uniform random \\ \hline
    \ac{AP} height & $5~\text{m}$ & $5~\text{m}$ \\ \hline
    $P_{\mathrm{max}}$ for \ac{AP} & $33~\text{dBm}$ & $30~\text{dBm}$ \\ \hline
    Number of \acp{UE} & $10$ & $10$\\ \hline
    \ac{UE} height & $1.7~\text{m}$ & $1.7~\text{m}$ \\ \hline
    \ac{UE} distribution & Uniform random & Uniform random \\ \hline
    Number of pilots & $\tau_3 = 8$ & $\tau_1 = 8$ \\ \hline
\end{tabular}
\label{table:network_topology}
\end{table}

The large-scale fading is modeled as $\beta = -30.5 - 36.7\log_{10} (d/1 \text{m})$ dB, where $d$ is the distance between an \ac{AP} and \ac{UE}. The spatially correlated shadow fading is generated using the \ac{SoS} method, as implemented in \cite{Pessoa2020}, with $200$ sinusoids, a standard deviation of $\sigma_{\text{sf}} = 8~\text{dB}$, and a decorrelation distance of $10~\text{m}$. The thermal noise variance is calculated as $\varsigma^2 = -174 + 10\log_{10}(W) + \text{NF}$~dBm. 

Both networks employ \ac{MR} linear precoding due to its low computational complexity and suitability for distributed implementations~\cite{Bjornson2020}. Additional simulation parameters are listed in \TabRef{table:parameters}.

\begin{table}[h!]
\centering
\caption{Simulation Parameters} 
\begin{tabular}{|l|c|}
    \hline
    \textbf{Parameter} & \textbf{Value} \\ 
    \hline \hline
    Bandwidth ($W$) & $20~\text{MHz}$ \\\hline
    Noise figure ($\text{NF}$) & $9~\text{dB}$ \\\hline
    Antenna spacing & $\lambda/2$ \\\hline
    Total number of pilots & $16$ \\\hline
    Coherence block length & $2000$ \\\hline
    Number of Monte Carlo realizations & $200$ \\\hline
\end{tabular}
\label{table:parameters}
\end{table}

\subsection{Zone Parameter Selection}

\begin{figure}[H]
    \centering
	\includegraphics[width=0.9\linewidth]{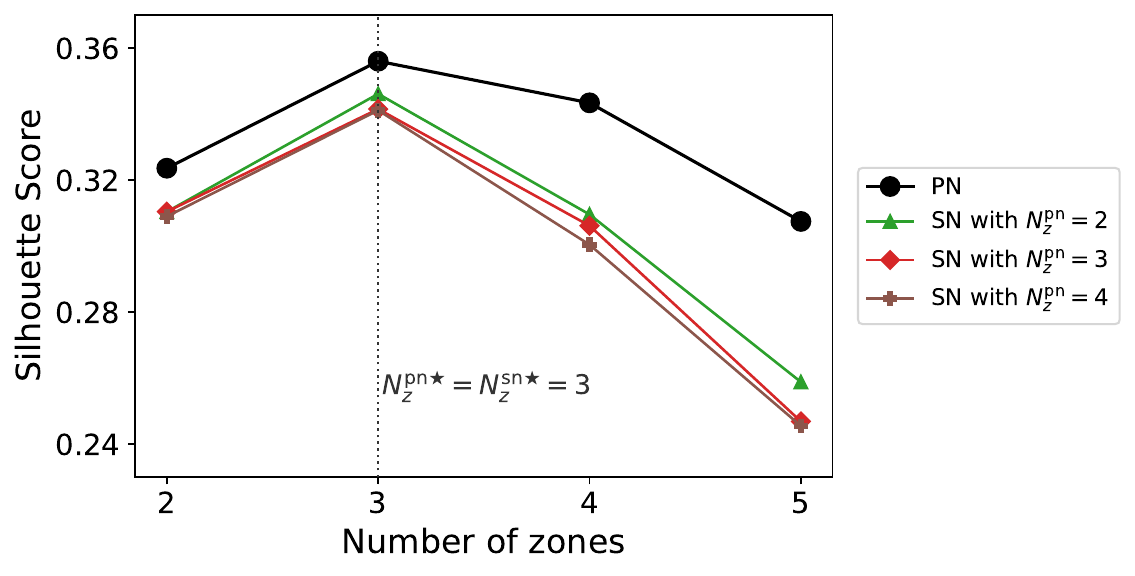}
	\caption{Zone separation quality (Silhouette Score) for \ac{PN} and \ac{SN} configurations.}
	\label{fig:silhouette}
\end{figure}

To select the zoning parameters, we evaluate the Silhouette Score across $N_z^{\pn} \in \{2,\ldots,5\}$ for the \ac{PN} and $N_z^{\sn} \in \{2,\ldots,5\}$ for the \ac{SN} with $N_z^{\pn} \in \{2, 3, 4\}$, as depicted in \FigRef{fig:silhouette}. For the \ac{PN}, the score peaks at $N_z^{\pn\star} = 3$ ($\bar{s} = 0.36$). For the \ac{SN}, the score consistently peaks at $N_z^{\sn\star} = 3$ across all evaluated values of $N_z^{\pn}$, with scores varying by less than $0.01$ between curves, indicating that the \ac{S-AP} risk classification is robust to the choice of $N_z^{\pn}$.

\begin{figure}[!h]
	\centering
	\includegraphics[width=0.8\linewidth]{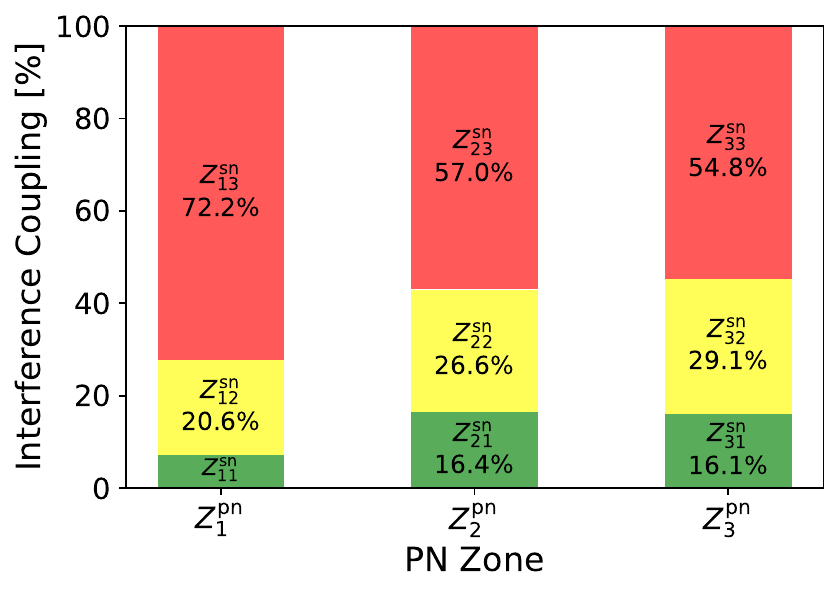}
    \caption{Interference coupling fractions under $N_z^{\pn\star} = N_z^{\sn\star} = 3$.}
	\label{fig:coupling_fracs}
\end{figure}

The validity of this stratification is further confirmed by the interference coupling fractions depicted in \FigRef{fig:coupling_fracs}. For a given \ac{PN} zone $Z_i^{\pn}$ and risk tier $Z_{it}^{\sn}$, the interference coupling fraction is defined as the share of the aggregate interference potential perceived by the \acp{P-UE} located in $Z_i^{\pn}$ that is contributed by the \acp{S-AP} in $Z_{it}^{\sn}$. For $N_z^{\pn\star} = N_z^{\sn\star} = 3$, high-risk \acp{S-AP} ($t = N_z^{\sn}$) account for at least $55\%$ of the coupling in every \ac{PN} zone, reaching $72.2\%$ in $Z_1^{\pn}$, while low-risk nodes ($t = 1$) contribute no more than $16\%$.

\begin{figure}[h]
	\centering	
    \includegraphics[width=0.9\linewidth]{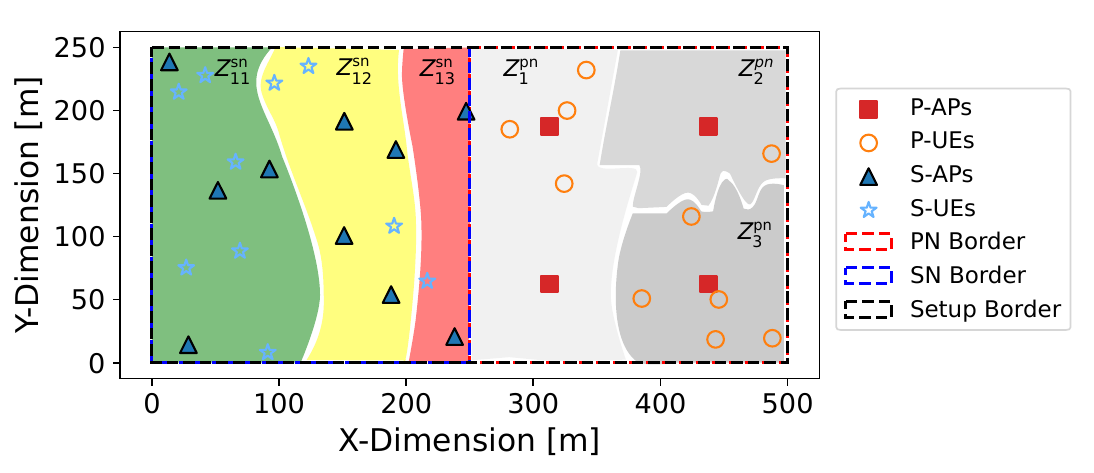}
	\caption{Coexistence scenario layout under Scenario~2 ($N_z^{\pn} = N_z^{\sn} = 3$): \ac{PN} zone partitioning and \ac{S-AP} risk classification with respect to $Z_1^{\pn}$ into low- (green), medium- (yellow), and high-risk (red) tiers.}
	\label{fig:scenario_layout_3zones}
\end{figure}

Thus, two configurations are evaluated: Scenario~1 with $N_z^{\pn} = 1$ and $N_z^{\sn} = 3$, treating the \ac{PN} as a single interference region (illustrated in \FigRef{fig:scenario_layout}), and Scenario~2 with $N_z^{\pn} = N_z^{\sn} = 3$, partitioning the \ac{PN} into three spatially distinct zones. \FigRef{fig:scenario_layout_3zones} illustrates the resulting \ac{PN} partitioning together with the \ac{S-AP} classification with respect to $Z_1^{\pn}$, the most exposed zone. The classification for the remaining zones follows analogously. Scenario~2 combines the finer partitioning with a more selective restriction and per-zone thresholds, both enabled by the zone-level evaluation.

\subsection{Power Allocation Schemes}

We consider three power allocation schemes. The first is a uniform baseline where each \ac{AP} distributes its total power equally among its served \acp{UE}, serving as a reference for uncoordinated coexistence. As a benchmark, the scheme proposed in~\cite{Shaik2024} constrains the transmit power allocated to the $\idxsue$-th \ac{S-UE} by the $\idxsap$-th \ac{S-AP} as

\begin{equation}
    \rho_{\idxsue \idxsap} = \min \left\{ \frac{P_{\max, \idxsap}}{\numSUE}, \frac{I_T}{\Gamma_1}, \dots, \frac{I_T}{\Gamma_{\numPUE}} \right\},
    \label{eq:power_allocation}
\end{equation}

\noindent where $I_T$ denotes the interference threshold and $\Gamma_{j}$ is the aggregate channel coupling for the $\idxpue$-th \ac{P-UE}, defined as 

\begin{equation}
    \Gamma_{\idxpue} =\sum_{\idxsap=1}^{\numSAP} \sum_{\idxsue=1}^{\numSUE} \mathbb{E}\{|\chSP_{\idxpue \idxsap}^{\mathrm{H}} \mathbf{w}_{\idxsue \idxsap}^{\mathrm{sn}}|^2\},
\end{equation}
\noindent which captures the expected aggregate interference leakage. While this approach ensures that the interference at each \ac{P-UE} strictly adheres to $I_T$, it enforces a spatially-blind uniform power reduction across all \acp{S-AP}, unnecessarily penalizing nodes with low interference potential.

The proposed scheme addresses this limitation by applying \eqref{eq:power_allocation} selectively per \ac{PN} zone $Z_i^{\pn}$, exclusively to \acp{S-AP} classified above a given risk threshold and evaluated solely against the \acp{P-UE} in $Z_i^{\pn}$, i.e., 

\begin{equation}
    \rho_{\idxsue \idxsap} = \min \left\{ \frac{P_{\max, \idxsap}}{\numSUE},\; \frac{I_{T,i}}{\max_{\idxpue \in \mathcal{J}_i}{\Gamma_{\idxpue}}} \right\},
    \label{eq:zone_constraint}
\end{equation}

\noindent where $I_{T, i}$ is the interference threshold for \ac{PN} zone $Z_i^{\pn}$ and $\mathcal{J}_i = \{\idxpue : \mathbf{p}_{\idxpue} \in Z_i^{\pn}\}$ denotes the set of \acp{P-UE} located in $Z_i^{\pn}$, so that the constraint is dictated by the most vulnerable \ac{P-UE} in the zone. If an \ac{S-AP} is classified above the risk threshold in multiple \ac{PN} zones, \eqref{eq:zone_constraint} is applied independently per zone and the minimum resulting power is retained, reflecting the cumulative interference threat posed by that node.

\subsection{Performance Evaluation}

\begin{figure}[h!]
	\centering	
	\subfloat[Primary network.]{\label{fig:cdf_pn}\includegraphics[width=0.88\linewidth]{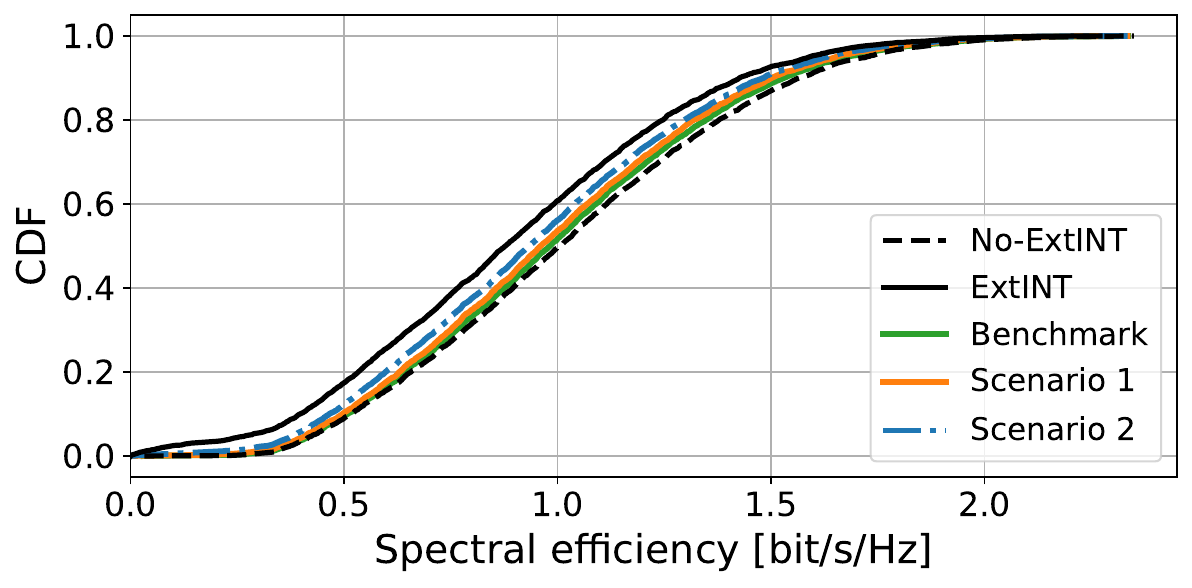}} \\
	\subfloat[Secondary network.]{\label{fig:cdf_sn}\includegraphics[width=0.88\linewidth]{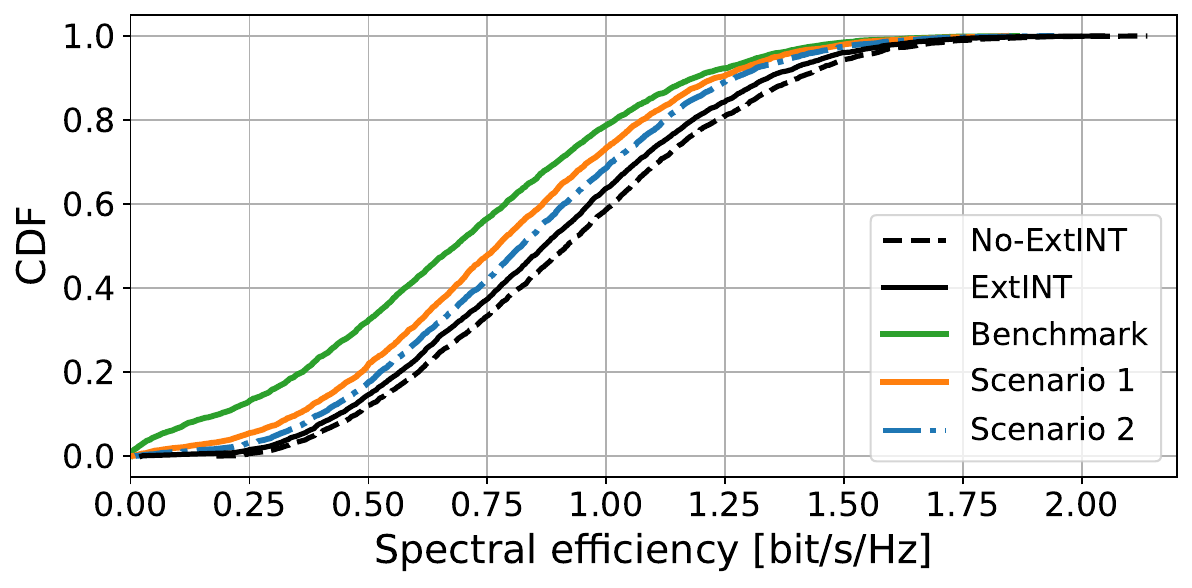}}
	\caption{Per-user SE CDF for the \ac{PN} (a) and \ac{SN} (b).}
	\label{fig:cdf}
\end{figure}

\FigRef{fig:cdf} presents the per-user \ac{SE} \ac{CDF} for the \ac{PN} and \ac{SN}. The No-ExtINT curve corresponds to the case without external interference, while ExtINT represents uncoordinated coexistence under uniform power allocation. These two curves thus delimit the \ac{PN} \ac{SE} range achievable by any coexistence scheme. Activating the \ac{SN} causes a noticeable left shift in the \ac{PN} \ac{SE}, confirming its sensitivity to cross-network interference. The benchmark scheme \eqref{eq:power_allocation}, evaluated at $I_T/\varsigma^2 = 4~\text{dB}$, closely restores \ac{PN} performance to the No-ExtINT baseline, but severely degrades \ac{SN} \ac{SE} due to the spatially-blind power reduction imposed on all \acp{S-AP}.

Under Scenario~1 $(N_z^{\pn} = 1,\, N_z^{\sn} = 3)$, \eqref{eq:zone_constraint} is applied to medium- and high-risk \acp{S-AP} with $I_{T,1}/\varsigma^2 = 4~\text{dB}$, i.e., the same threshold adopted by the benchmark, while low-risk nodes operate at full power.  The resulting \ac{SN} \ac{SE} gain is therefore attributable solely to the selective application of the constraint, and \ac{PN} protection remains near the No-ExtINT baseline. Under Scenario~2 $(N_z^{\pn} = N_z^{\sn} = 3)$, the constraint is enforced independently per \ac{PN} zone and applied exclusively to high-risk \acp{S-AP}, with thresholds $I_{T,1}/\varsigma^2 = 4~\text{dB}$, $I_{T,2}/\varsigma^2 = 5~\text{dB}$, and $I_{T,3}/\varsigma^2 = 6~\text{dB}$, yielding a further \ac{SN} \ac{SE} improvement enabled jointly by the finer partitioning, the stricter risk tier, and the relaxed limits in the less exposed zones.

As shown in \TabRef{tab:percentiles}, Scenario~2 increases the \ac{SN} \ac{SE} at the 10th percentile from $0.187$ to $0.400$~bit/s/Hz relative to the benchmark, a gain of $114\%$, at the cost of a $10.0\%$ reduction in \ac{PN} \ac{SE} relative to the No-ExtINT baseline ($0.466$ vs.\ $0.518$~bit/s/Hz), demonstrating that zone-aware selective mitigation effectively decouples \ac{PN} protection from \ac{SN} performance degradation.

\begin{table}[h]
\centering
\caption{\ac{SE} 10th and 50th percentiles for all power allocation schemes.} 
\begin{tabular}{|l|c|c|c|}
    \hline
    & & & \\[-8pt] 
    \multirow{2}{*}{\textbf{Scheme}} & \textbf{PN 10th} & \textbf{SN 10th} & \textbf{SN 50th} \\
    & [bit/s/Hz] & [bit/s/Hz] & [bit/s/Hz] \\
    \hline\hline
    No-ExtINT & $0.518$ & $0.476$ & $0.918$  \\ \hline
    ExtINT & $0.396$ & $0.438$ & $0.869$     \\ \hline
    Benchmark & $0.508$ & $0.187$ & $0.683$  \\ \hline
    Scenario 1 & $0.496$ & $0.351$ & $0.771$ \\ \hline
    Scenario 2 & $0.466$ & $0.400$ & $0.822$ \\ \hline
\end{tabular}
\label{tab:percentiles}
\end{table}

Since \TabRef{tab:percentiles} reports the \ac{PN} \ac{SE} aggregated over all \acp{P-UE}, the zone-specific thresholds adopted in Scenario~2 raise the question of whether protection is distributed evenly across the \ac{PN} zones. \TabRef{tab:pn_zones} addresses this by reporting the \ac{PN} 10th-percentile \ac{SE} separately for each zone. The No-ExtINT, ExtINT, and benchmark schemes, all blind to the \ac{PN} partitioning, establish the baseline profile induced by the network geometry alone, so that any deviation of Scenario~2 from this profile is attributable to the zone-selective mitigation.

\begin{table}[h!]
\centering
\caption{10th percentile of the \ac{PN} \ac{SE} per zone.}
\begin{tabular}{|l|c|c|c|}
    \hline
    & & & \\[-7pt] 
    \multirow{2}{*}{\textbf{Scheme}} & \textbf{$Z_1^{\pn}$} & \textbf{$Z_2^{\pn}$} & \textbf{$Z_3^{\pn}$} \\
    & [bit/s/Hz] & [bit/s/Hz] & [bit/s/Hz] \\
    \hline\hline
    No-ExtINT & $0.502$ & $0.524$ & $0.521$  \\ \hline
    ExtINT & $0.235$ & $0.490$ & $0.488$     \\ \hline
    Benchmark & $0.488$ & $0.521$ & $0.520$  \\ \hline
    Scenario 2 & $0.413$ & $0.507$ & $0.509$ \\ \hline
\end{tabular}
\label{tab:pn_zones}
\end{table}

The results show that the interference concentrates almost entirely in $Z_1^{\pn}$, the most exposed zone, whose 10th-percentile \ac{SE} collapses to $0.235$~bit/s/Hz under uncoordinated coexistence (ExtINT), while $Z_2^{\pn}$ and $Z_3^{\pn}$ remain above $0.48$~bit/s/Hz. Scenario~2 restores $Z_1^{\pn}$ to $0.413$~bit/s/Hz, a $76\%$ gain over uncoordinated coexistence, while leaving the already-protected zones essentially unchanged. Note that $Z_1^{\pn}$ is also the weakest zone under No-ExtINT, reflecting the \ac{PN} geometry rather than the mitigation itself. Protection is thus applied where it is most needed, consistent with the differentiated thresholds.

\section{Conclusions}\label{section_conclusion}

This work proposed a zoning-based interference management framework for underlay spectrum sharing in cell-free networks. By classifying \acp{S-AP} according to their interference potential and applying selective power control at the zone level, the framework reduces coordination overhead while preserving \ac{SN} utility. Simulation results confirmed that the proposed strategy enables proper coexistence with limited impact on the \ac{PN}, achieving a $114\%$ gain in the \ac{SN} 10th-percentile \ac{SE} over the benchmark scheme, at the cost of a $10\%$ reduction in \ac{PN} \ac{SE} relative to the No-ExtINT baseline, demonstrating that zone-based interference abstraction provides an efficient trade-off between coexistence performance and coordination complexity.

\end{document}